\documentstyle[emulateapj,psfig]{article}

\makeatletter

\newenvironment{inlinefigure}{%
\def\@captype{figure}%
\noindent\begin{minipage}{0.999\linewidth}\begin{center}}
{\end{center}\end{minipage}\smallskip}
\makeatother

\lefthead{Cowie, Barger, \& Kneib}

\begin{document}
\title{Faint Submillimeter Counts from Deep 850 Micron
Observations of the Lensing Clusters A370, A851, and A2390}
\author{L.\,L.\ Cowie,$\!$\altaffilmark{1}
A.\,J.\ Barger,$\!$\altaffilmark{2,3,1}
J.-P.\ Kneib$\!$\altaffilmark{4}
}

\altaffiltext{1}{Institute for Astronomy, University of Hawaii,
2680 Woodlawn Drive, Honolulu, Hawaii 96822}
\altaffiltext{2}{Department of Astronomy, University of Wisconsin-Madison,
475 North Charter Street, Madison, WI 53706}
\altaffiltext{3}{Department of Physics and Astronomy,
University of Hawaii, 2505 Correa Road, Honolulu, HI 96822}
\altaffiltext{4}{Observatoire Midi-Pyr{\'e}n{\'e}es,
14 Avenue E. Belin, 31400 Toulouse, France}

\slugcomment{Astronomical Journal in press}

\begin{abstract}

We present deep 850~$\mu$m maps of three massive lensing clusters,
A370, A851, and A2390, with well-constrained mass models. 
Our cluster exposure times are more than 2 to 5 times longer than
any other published cluster field observations. We catalog the sources 
and determine the submillimeter number counts. The counts are best 
determined in the 0.3 to 2~mJy range where the areas are large enough 
to provide a significant sample. At 0.3~mJy the cumulative counts
are $3.3_{1.3}^{6.3}\times 10^4$~deg$^{-2}$, where the upper and
lower bounds are the 90\% confidence range. The surface density at these faint
count limits enters the realm of significant overlap with other galaxy
populations.The corresponding percentage 
of the extragalactic background light (EBL) 
residing in this flux range is about $45-65$\%, depending on the EBL
measurement used. Given that 
$20-30$\% of the EBL is resolved at flux densities between 2 and 
10~mJy, most of the submillimeter EBL is arising 
in sources above 0.3~mJy. We also performed a noise analysis to 
obtain an independent estimate of the counts. The upper bounds on 
the counts determined from the noise analysis closely match 
the upper limits obtained from the direct counts.
The differential counts from this and other surveys 
can reasonably be described by the parameterization 
$n(S)=3\times 10^4$~deg$^{-2}$~mJy$^{-1}/(0.7 + S^{3.0})$
with $S$ in mJy, which also integrates to match the EBL.

\end{abstract}


\section{Introduction}
\label{secintro}

The cumulative emission from all sources lying beyond the Galaxy, the
extragalactic background light (EBL), provides important constraints on
the integrated star formation and accretion histories of the Universe. 
{\it COBE} measurements of the EBL at far-infrared (FIR) and 
submillimeter wavelengths (e.g., \markcite{puget96}Puget et al.\ 1996; 
\markcite{fixsen98}Fixsen et al.\ 1998) 
indicate that the total radiated emission 
that is absorbed by dust and gas and then reradiated into the 
FIR/submillimeter is comparable to the total measured optical EBL. 
Deep submillimeter surveys with SCUBA 
(\markcite{holland99}Holland et al.\ 1999)
on the James Clerk 
Maxwell Telescope\altaffilmark{5}\altaffiltext{5}{The JCMT is operated by
the Joint Astronomy Centre on behalf of the parent organizations
the Particle Physics and Astronomy Research Council in the United
Kingdom, the National Research Council of Canada, and The Netherlands
Organization for Scientific Research.}
have uncovered the
brighter obscured sources which give rise to a substantial part of the 
FIR/submillimeter EBL.
The properties of these sources are similar to the properties 
of the most luminous systems observed locally, the ultraluminous 
infrared galaxies (ULIGs, \markcite{sanders96}Sanders \& Mirabel 1996).

Blank field SCUBA surveys have resolved sources over the 
$2-10$~mJy range that account for $20-30$\% of the 850~$\mu$m 
EBL (e.g., \markcite{barger98}Barger et al.\ 1998;
\markcite{hughes98}Hughes et al.\ 1998; 
\markcite{eales99}Eales et al.\ 1999, 2001;
\markcite{bcs99}Barger, Cowie, \& Sanders 1999a;
\markcite{scott01}Scott et al.\ 2002;
\markcite{borys02}Borys et al.\ 2002;
\markcite{webb02}Webb et al.\ 2002). 
\markcite{bcs99}Barger et al.\ (1999a) found that their
cumulative source counts per square degree were well
described by the power-law parameterization of the differential
counts

$$n(S)=N_0/(a+S^{\alpha})$$ 

\noindent
with $S$ in mJy, $\alpha=3.2$, and
$N_0=3.0\times 10^4$~deg$^{-2}$~mJy$^{-1}$.
Assuming only that their parameterization provided
an appropriately smooth continuation to fluxes below 2~mJy,
\markcite{bcs99}Barger et al.\ (1999a) constrained their fit to 
match the EBL with $a\sim 0.5$. Using this empirical parameterization, 
they predicted that very approximately 60\% of the EBL at 850~$\mu$m 
should be resolved into discrete sources between 0.5 and 2.0~mJy.

Blank field SCUBA surveys cannot reach the required sensitivities 
to directly detect the dominant population of $<2$~mJy
extragalactic sources due to confusion noise resulting
from the coarse resolution of SCUBA (e.g., \markcite{h01}Hogg 2001). 
Thus, in order to search for 
this population with SCUBA, one must observe fields with massive 
cluster lenses to take advantage of both gravitational amplification 
by the lens and reduced confusion noise.
\markcite{smail97}Smail, Ivison, \& Blain (1997) 
and \markcite{smail98}Smail et al.\ (1998) pioneered this
method, making the first 
SCUBA observations of cluster lenses. They studied seven
clusters with well-constrained lens models where it is possible to  
correct the observed source fluxes for lens amplification.
Their survey was designed to detect the brightest submillimeter 
sources in relatively short integration times, and hence it is 
quite shallow with $3\sigma$ flux limits around 6~mJy. 

Despite the shallowness of the survey,
\markcite{blain99}Blain et al.\ (1999)
tried to determine the number counts at and below 1~mJy,
arguing that the small regions of high amplification
(around a factor of 6 at 1~mJy and a factor of 24
at 0.25~mJy) could be used to probe these faint fluxes. 
(A later analysis of another cluster sample of similar depth
by \markcite{chapman02}Chapman et al.\ (2002) did not extend 
the counts below 1~mJy.)
However, direct inversion at these amplification levels
is complicated because redshift uncertainties and small positional 
changes can have large effects on the amplification.
In SCUBA surveys the positions of the 
submillimeter sources are relatively poorly determined and the 
redshifts are often unknown.
While these effects have comparatively little effect
on the shape of the determined counts, they do effectively
limit the flux level to which the counts can be considered
to be determined.
We shall discuss this point further in \S~4.

The sub-mJy counts are best addressed by obtaining
much deeper images, since at lower amplifications there
is considerably less redshift and positional sensitivity. 
At typical amplifications of 1 to 4, the counts can be 
robustly investigated down to a few tenths of a mJy
from images with $3\sigma$ limits of 1.5 to 2~mJy.
Because of this key point, and because the observed areas at 
the sub-mJy fluxes of interest rise rapidly with deeper 
obervations, it is a natural and important progression to pursue 
the faint submillimeter counts with much longer integrations on 
lensing cluster fields. This is the subject of the present paper.

\section{Observations and Sample}
\label{secdata}

SCUBA jiggle map observations were taken in mostly excellent
observing conditions during runs in 1999 August, September,
and November; 2000 November and December; and
2001 January, March, and June.
The maps were dithered to prevent any regions of the sky from
repeatedly falling on bad bolometers. The chop throw was
fixed at a position angle of 90~deg so that the negative beams
would appear $45''$ on either side east-west of the positive beam.
Regular ``skydips'' (\markcite{manual}Lightfoot et al.\ 1998)
were obtained to measure the zenith atmospheric opacities at
450 and 850~$\mu$m, and the 225~GHz sky
opacity was monitored at all times to check for sky stability.
Pointing checks were performed every hour during the observations 
on the blazars 0106+013, 0215+015, 0221+067, 0336-019, 
0420-014, 0917+449, 0923+392, 2145+067, or 2251+158.

The data were reduced using
the dedicated SCUBA User Reduction Facility
(SURF, \markcite{surf}Jenness \& Lightfoot 1998).
Due to the variation in the density of bolometer samples across
the maps, there is a rapid increase in the noise levels at the
very edges, so the low exposure edges were clipped.
The data were calibrated using jiggle maps of the primary
calibration source Mars or the secondary calibration sources
CRL618, CRL2688, or OH231.8+4.2.
The routines produce a noise-weighted exposure time at each
pixel, which we shall denote as $t$.
Submillimeter fluxes were measured using beam-weighted 
routines that include both the positive and negative portions of
the beam profile to provide the optimal extraction.

The SURF reduction routines arbitrarily normalize all the data
maps in a reduction sequence to the central pixel of the first
map; thus, the noise levels in a combined image are
determined relative to the quality of the central pixel in the
first map. In order to determine the absolute noise levels of
our maps, we first eliminated the $\gtrsim 3\sigma$ real sources
in each field by subtracting an appropriately normalized version
of the beam profile. We then iteratively adjusted the noise
normalization until the dispersion of the signal-to-noise
ratio measured at random positions became $\sim 1$. This noise
estimate includes both fainter sources and correlated noise;
hereafter we refer to it as ``uncleaned noise''.

We also constructed ``true noise'' maps in which all the
sources have been cleaned out so that the maps  
contain only sky and bolometer noise. 
To do this, we divided the data for each cluster into two halves 
and then combined the jiggle maps for each of the two halves 
separately, keeping only the first jiggle map in the reduction 
sequence the same in both halves, since the data maps are
normalized to the central pixel of the first map. We then
subtracted the two halves from one another. Since the faint
sources are at fixed positions in the maps, the subtraction
effectively removes them from the maps, resulting in true noise
maps for each cluster field. The true noise maps were then scaled
to the actual maps by multiplying each pixel by the factor
$(t_1+t_2)/\sqrt{(t_1\times t_2})$, where $t_1$ and $t_2$ are 
the weighted exposure times in the two half images.

\begin{figure*}[tb]
\centerline{\psfig{figure=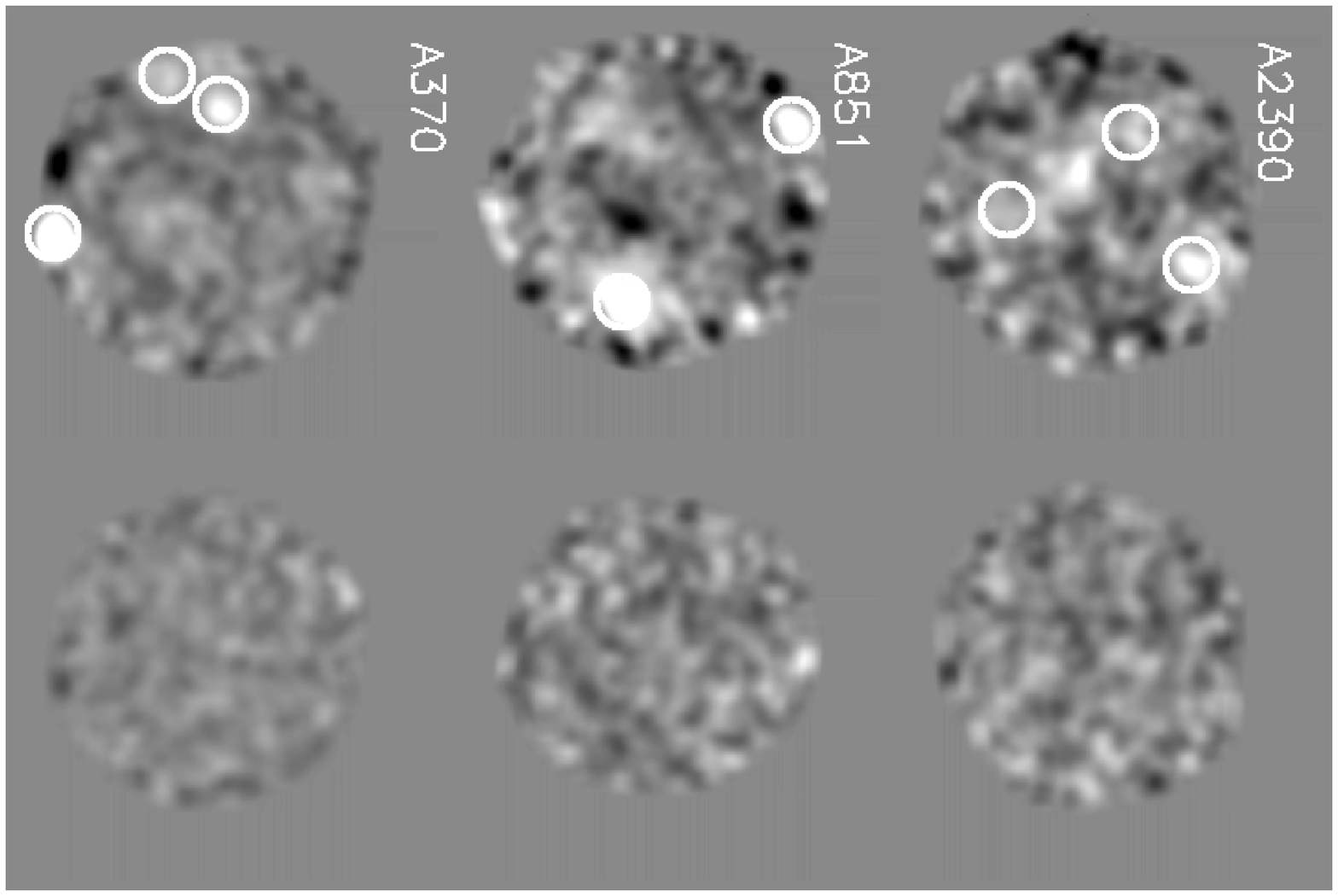,angle=90}}
\figurenum{1}
\figcaption[]{
SCUBA maps of the A370, A851, and A2390 cluster fields (left) and 
their corresponding true noise maps (right). The $4\sigma$ sources
listed in Table~3 are circled in the cluster images in the 
left-hand panels. 
\label{fig1}
}
\end{figure*}

The three cluster field centers, total SCUBA exposure times 
(based on the number of integrations), median optical depths, 
areas, and 850~$\mu$m sensitivities ($1\sigma$) for the true 
noise maps and the uncleaned noise maps are given in Table~\ref{tab1}.
Our cluster exposure times are more than 2 to 5 times longer 
than any other published cluster field observations
(\markcite{smail97}Smail et al.\ 1997, 1998; 
\markcite{chapman02}Chapman et al.\ 2002).
Our three cluster field SCUBA maps and their corresponding true 
noise maps are shown in Fig.~\ref{fig1}.

\section{Source Catalogs and Identifications}

In contrast to the situation in the optical and near infrared, we do not expect 
there to be much contamination of the background submillimeter number counts
by cluster members, which lie at too low redshifts to be
strong submillimeter sources. However, there may be submillimeter
sources produced by the cooling flow regions in the clusters
and the associated strong radio sources in some of the
central cluster galaxies (\markcite{edge00}Edge et al.\ 1999).
In particular, \markcite{edge99}Edge et al.\ report a strong 
source associated with the cD galaxy in A2390.
We therefore searched the images for sources of this type.
Table~\ref{tab2} gives the source number,
measured positions (columns 2 and 3),
measured submillimeter fluxes and uncertainties (column 4),
and measured signal-to-noise ratios (column 5)
for the D cluster galaxies in A370 and A851
and the cD galaxy in A2390. We confirm the detection of the
cD galaxy in A2390 and also find a source associated with
the southern D galaxy in A370. These two sources were removed
from the maps by subtracting a normalized image of the beam.

To generate the catalog of the background submillimeter galaxies,
we next scanned the three SCUBA maps at an array of positions
separated by $3''$ to make a preliminary selection of candidate
sources. Using peaked-up positions, we then located the
$\ge3\sigma$ sources. We subtracted the brighter sources from
the maps before extracting the fainter sources in order
to avoid contamination by the large and complex beam pattern.
Table~\ref{tab3} summarizes the information for the background 
sources in the same format as Table~\ref{tab2}.
Only sources with weighted exposure times greater than a tenth 
of the maximum in the image are included, giving a total sample 
of 15 sources. For the purpose of the counts analysis,
the $3\sigma$ selection is reasonable; however, considerable caution should
be used in interpreting sources at this low significance level
since there may be substantial uncertainties in both position and
flux. Our simulations (see \S~\ref{seccounts}) suggest that 
interpretation of individual sources and attempts to identify 
counterparts are best restricted to at least the $\ge 4\sigma$ subsample
(see also \markcite{scott02}Scott et al.\ 2002).
In the last three columns of Table~\ref{tab3}
we give the previous submillimeter flux measurements from
\markcite{smail97}Smail et al.\ (1997, 1998) for the sources
detected in their shallow survey, along with
the RA and Dec offsets between our measured source
positions and theirs. In general the fluxes agree to
within the noise uncertainty. However, the offsets in position
range up to $4''$ for the faintest overlapping source. 

%
% Figure 2
%

\begin{inlinefigure}
\psfig{figure=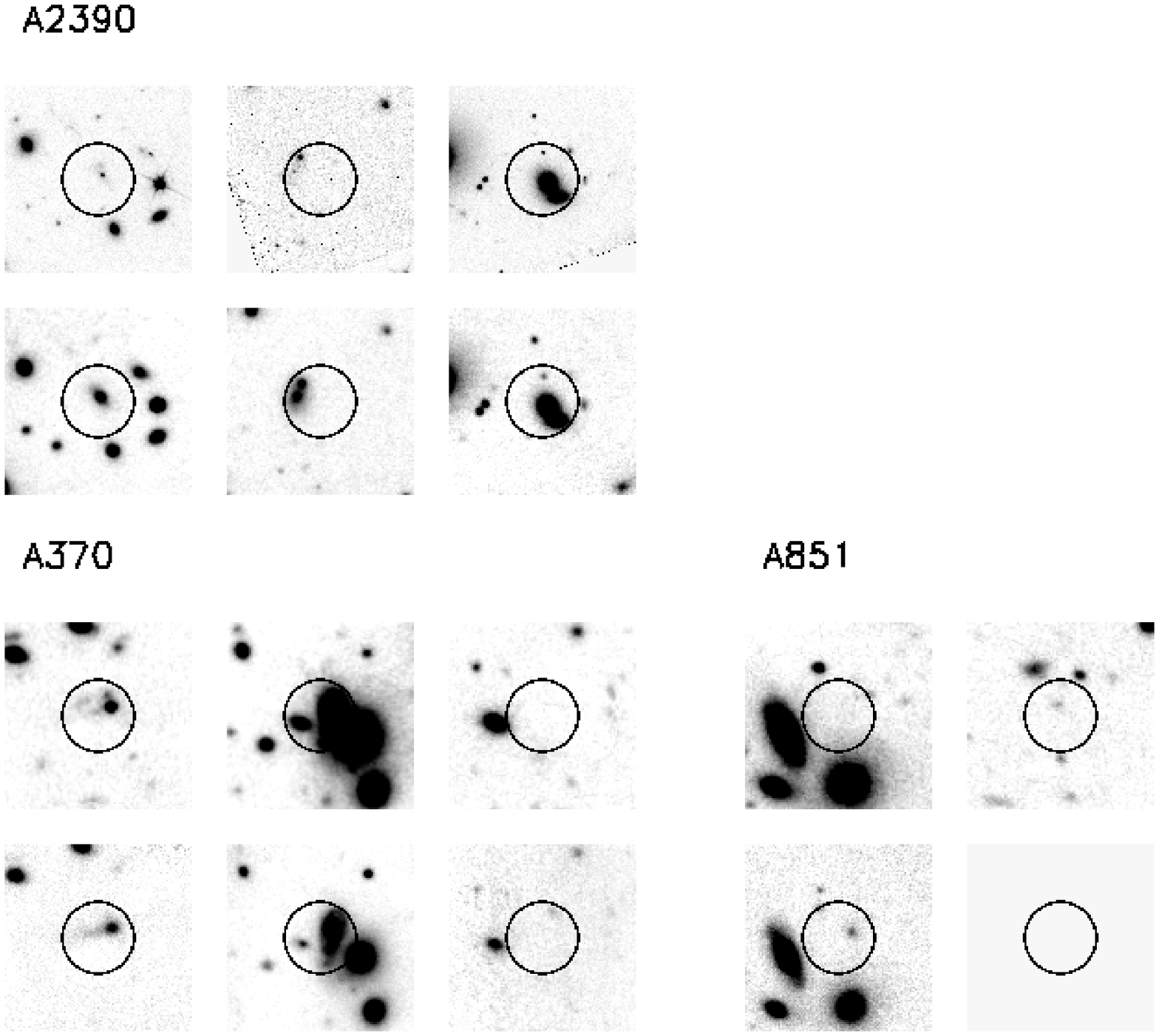,angle=90,width=3.5in}
\vspace{6pt}
\figurenum{2}
\caption{$K$ (lower panel) and $I$ (upper panel) thumbnail images 
for the $4\sigma$ sources in Table 1.~The ordering within
each cluster (left = submillimeter bright to right = submillimeter faint) 
follows that of Table~3. The panels are $19''$ on a side, centered on
the nominal SCUBA position. The circles are of radius $3.5''$, which
is representative of the positional uncertainty. North is up and East is 
to the right.
\label{fig2}
}
\addtolength{\baselineskip}{10pt}
\end{inlinefigure}

Efforts to spectroscopically identify the counterparts
to the SCUBA sources have proved difficult due to the optically
faint nature of the majority of the counterparts; however,
high-quality optical images, extremely deep radio maps,
and follow-up optical spectroscopy suggest that the bulk of
the submillimeter sources lie in the redshift range $z=1-3$
(e.g., \markcite{smail98}Smail et al.\ 1998;
\markcite{barger99b}Barger et al.\ 1999b;
\markcite{smail00}Smail et al.\ 2000;
\markcite{cy00}Carilli \& Yun\ 2000;
\markcite{bcr00}Barger, Cowie, \& Richards 2000; see, however,
\markcite{lilly99}Lilly et al.\ 1999).

In Fig.~2 we show $I$ and $K$-band thumbnail images of the eight $4\sigma$
sources in the sample, according to cluster. The image is taken from
E.M.~Hedrick, A.J.~Barger, \& J.-P.~Kneib, in preparation, where the data
reduction and photometry of the sources are described. The two brightest
submillimeter sources in A370 both have spectroscopic redshifts
($z=2.80$ from \markcite{ivison98}Ivison et al.\ 1998 and
$z=1.06$ from \markcite{barger99b}Barger et al.\ 1999b and
\markcite{soucail99}Soucail et al.\ 1999), which we give
in column~7 of Table~2. The other two sources with previous
discussions in the literature are the brightest submillimeter source
in A851, believed to be associated with the
extremely red object (ERO) in the error circle
(\markcite{i00}Ivison et al.\ 2000), and the second brightest
submillimeter source
in the A2390 sample. The most likely counterpart to the
latter is thought to be the ERO at $z=1.02$, which is to the
West of the thumbnail center in Fig.~2.
This source (designated K3 in \markcite{barger99b}Barger et al.\ 1999b
where the redshift was given) is also a weak ISOCAM source 
(\markcite{l99}Lemonon et al.\ 1999).
However, the large separation from the Smail et al.\ position
($6.3''$) made the identification suspect. With the new position
the source now lies within the submillimeter error circle, and
we now consider this identification to be highly probable.
We give this redshift in Table~2. 

The remaining sources are most likely associated with EROs in
their error circles. In particular, the brightest A2390 submillimeter
source is associated with a very luminous ERO, which is also a
weak ISOCAM source (\markcite{l99}Lemonon et al.\ 1999). We do not 
know of a spectroscopic identification for this source. The weakest of
the three $4\sigma$ A2390 sources lies near a bright spectroscopically
confirmed cluster member. It appears unlikely this is
the correct identification of the source, which may be associated
with one of the fainter galaxies in this area. The effects
of excluding this source from our subsequent analysis are small.
A more extensive discussion of the submillimeter properties
of the optical and near-infrared selected galaxies in these fields
may be found in E.M.~Hedrick, A.J.~Barger, \& J.-P.~Kneib, 
in preparation.

The observed sources directly determine the fraction of the
EBL which has been resolved, since the average sky brightness
is conserved by the lensing process. This is by far the
most robust quantity which can be derived from the observations
(\markcite{chapman02}Chapman et al.\ 2002). The average EBL contribution of
the three fields is 17.3~Jy~deg$^{-2}$, and the breakdown
by field is 22.7~Jy~deg$^{-2}$ (A370), 17.4~Jy~deg$^{-2}$ (A851), 
and 11.8~Jy~deg$^{-2}$ (A2390), with
the differences reflecting the different depths of the images
and the statistical uncertainties from the small numbers of
sources involved. The average over the three fields is
between 40\% and 56\% of the total EBL, depending on whether
we adopt the EBL value of 44~Jy~deg$^{-2}$ (Fixsen
et al.\ 1998) or 31~Jy~deg$^{-2}$ (Puget et al.\ 1996).
However, detailed models are required to determine the
flux ranges which give rise to this contribution.
The contribution to the EBL from sources greater than 6~mJy
(roughly previous survey limits) is 7.8~Jy~deg$^{-2}$;
thus, the increased sensitivity roughly doubles the resolved 
fraction of the EBL in these clusters. 

\section{Direct Number Counts}
\label{seccounts}

For a blank field observation, the number counts are determined by 
dividing the number of significantly detected sources in a sample 
by the area over which the sources could be detected at that level. 
In the case of a lensed field, the sensitivity to a background 
source is dependent on the position and redshift of the background 
source and on the redshift of the gravitational lens.  
These lensing effects can be corrected for with a 
sufficiently detailed mass model for the lens.

We used the
{\sc LENSTOOL} models to determine the flux amplifications.
{\sc LENSTOOL} uses multiple-component mass distributions
that describe the extended potential well of the clusters and
their more massive individual member galaxies
(e.g., \markcite{kneib96}Kneib et al.\ 1996). The mass
distributions are derived from the positions of multiply-imaged
features identified in high-resolution optical images;
spectroscopic redshifts constrain the models. Details of the models 
for the A370, A851, and A2390 clusters can be
found in \markcite{kneib93}Kneib et al.\ (1993),
\markcite{seitz96}Seitz et al.\ (1996), and
Kneib et al., in preparation, respectively.
Using {\sc LENSTOOL} we mapped the background galaxies from
their observed positions back onto the source plane using known
redshifts, where available, or to an assumed source
redshift of $z=3$. We give the resulting amplifications in 
column~6 of Table~\ref{tab3}.

Although the amplification of a source
depends on both the redshift of the lens and the redshift of the
source, at the redshifts of our three cluster lenses
($z=0.37$ for A370, $z=0.41$ for A851, and $z=0.23$ for A2390),
the amplification varies only weakly with redshift for any source
beyond $z=1$ {\it which has a modest amplification}.
\markcite{blain99}Blain et al.\ (1999) estimated that as long
as the source redshifts are greater than one, the systematic
uncertainties (due to the uncertainties in both the redshift 
distribution of the detected sources and the mass models of 
the clusters) are less than 25\% and hence are comparable to 
the typical uncertainties in the absolute flux calibration of 
the SCUBA maps. 

However, sources with high amplifications have much 
larger uncertainties associated with positional uncertainty and 
redshift indeterminancy than sources with relatively low 
amplifications. For the high amplification
sources lying near critical lines, small changes in position
or redshift can result in very large variations in amplification.
In order to quantify this effect, we computed the maximum
and minimum amplifications in a $3''$ radius region surrounding
each source. For the sources where we do not 
have a refined position from an optical identification,
we give this range in brackets after the amplification in Table~3.
For sources with ``typical'' amplifications of 1 to 4 the effect
is generally small. However, the high amplification sources can
easily have order of magnitude amplification uncertainties associated 
with their positions (and also with their redshifts), 
and this must be allowed for in any analysis. 

Five of the sources in our sample (two in A370 and three in A2390) 
fall into this category. All lie at the faint end of the sample 
where we cannot easily use secondary constraints, such as the 
lack of multiple images, to provide limits on the amplifications. 
For two of the sources in A2390
(13 and 14 in Table~3) there are no sensible upper limits to
the amplifications, so we give only lower bounds.
In what follows we use the maximum and minimum amplifications 
given in Table~3 to determine how much the number counts can be 
changed by the amplification uncertainties. 

For the direct number counts, we need to know
the source plane areas for which background galaxies would fall 
within the SCUBA maps and be significantly detected. 
(Due to the expansion of the source plane, the source plane
areas are smaller than the areas of the SCUBA maps.) 
To do this, we created a grid of $z=3$ background sources at one 
arcsecond separations and used {\sc LENSTOOL} to produce the 
corresponding image plane maps with magnifications $m$. 
(For sources with known redshifts, we created the source plane 
grids at the known redshifts.)
Each grid point on the image plane corresponds to 
an area on the source plane of 1~arcsec$^2$. If the sensitivities
of the SCUBA maps were uniform, we could determine the source plane
area for a $3\sigma$ detection
at a given submillimeter flux by finding the number 
of points at each image plane grid point that satisfied the 
equation $3\sigma/mf(850~\mu$m) and then multiplying that number
by 1~arcsec$^2$. 

However, since the SCUBA maps become less 
sensitive towards the edges where the exposure times are less, 
the equation to be satisfied is instead
$3\sigma_{min}/(m\sqrt{t/t_{max}})<f(850~\mu$m),
where $\sigma_{min}$ is the minimum noise in the field and 
$t_{max}$ is the maximum weighted exposure time. The number
of points satisfying this equation can then be multiplied by 
1~arcsec$^2$.

%
% Figure 3
%

\begin{inlinefigure}
\psfig{figure=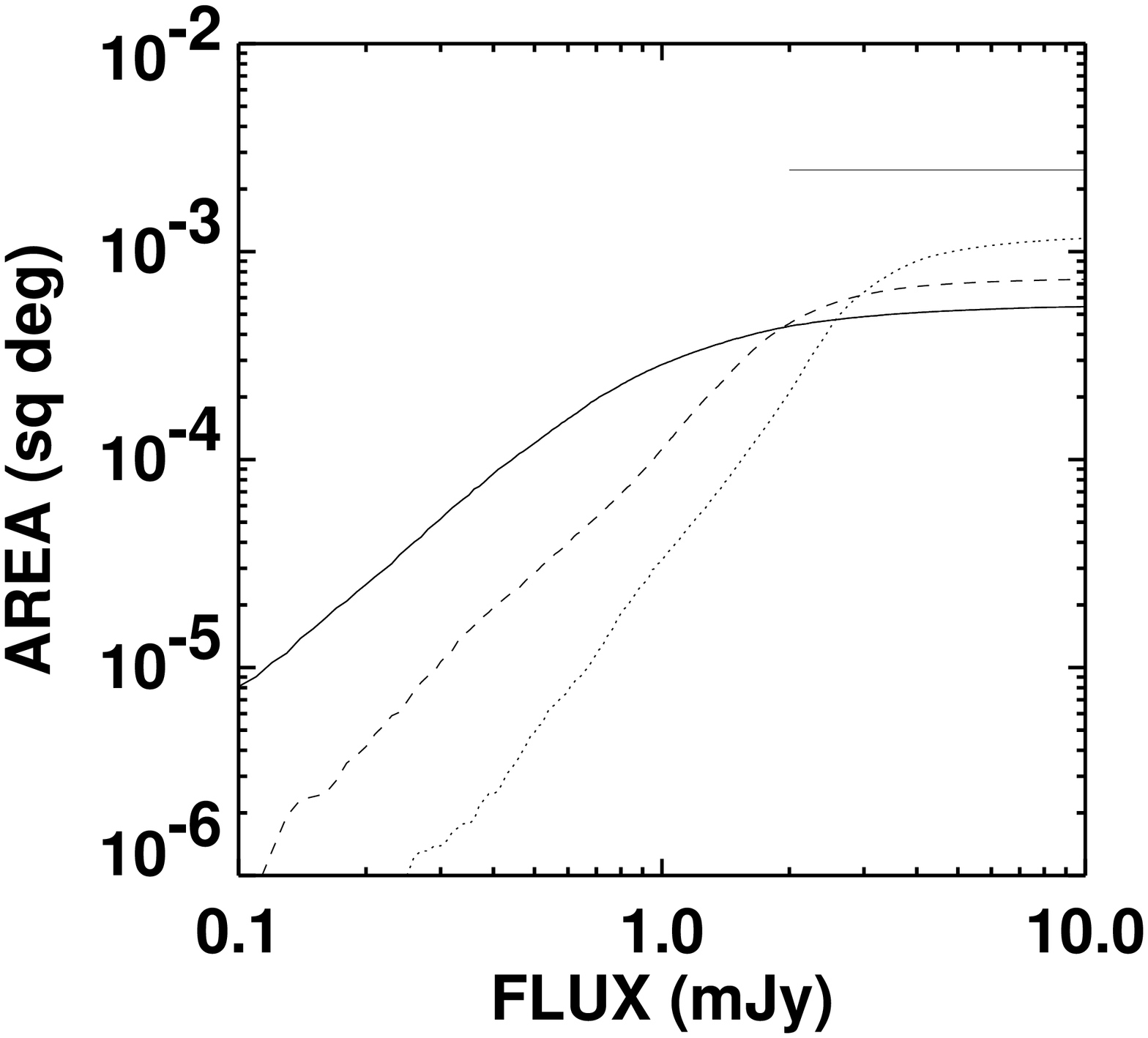,angle=90,width=3.5in}
\vspace{6pt}
\figurenum{3}
\caption{
Source plane areas, assuming source redshifts $z=3$,
over which a source of a given flux in the source plane
would be detected at the $3\sigma$ level in the image
plane. The three curves are for the A370
(solid), A851 (dotted), and A2390 (dashed) fields.
The horizontal line segment shows the SCUBA map area.
\label{fig3}
}
\addtolength{\baselineskip}{10pt}
\end{inlinefigure}

Figure~\ref{fig3} shows the areas over which a source with
a given flux in the source plane would be detected at the $3\sigma$ 
level in the image plane for each of the three cluster 
fields. The high magnification of the A370 cluster lens means
there is less source plane area at high submillimeter
fluxes in the A370 field but much more source plane area at low
submillimeter fluxes. The SCUBA data available on A370 are
deeper than the data on the other two clusters, which also
contributes to the sizeable differences in source plane areas 
seen in Fig.~\ref{fig3} at the fainter fluxes. 

\bigskip

%
% Figure 4
%

\begin{inlinefigure}
\psfig{figure=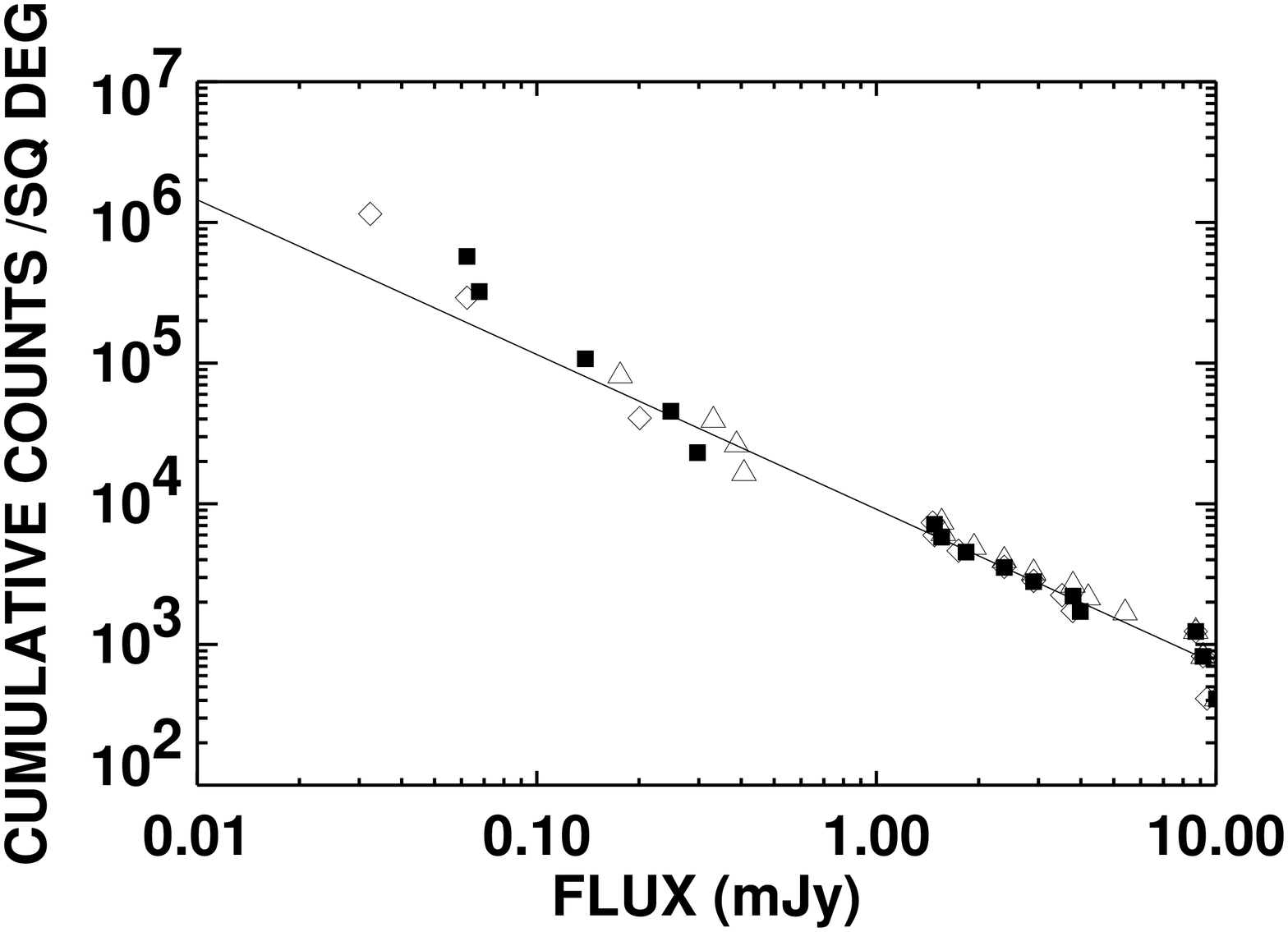,angle=90,width=3.5in}
\vspace{6pt}
\figurenum{4}
\caption{
Cumulative 850~$\mu$m source counts (solid squares).
The solid curve shows an area-weighted maximum likelihood
power-law fit to the data (solid squares) over the flux range from
0.1~mJy to 5~mJy. The slope is $-1.1$. 
The open triangles show the counts determined if we 
use the minimum amplifications of Table~3. The
open diamonds show the counts determined if we use the maximum 
amplifications.
\label{fig4}
}
\addtolength{\baselineskip}{10pt}
\end{inlinefigure}

We can now determine our raw cumulative source counts by summing the
inverse areas of all the sources in the three fields brighter
than flux $S$. We present these raw cumulative 850~$\mu$m counts
per square degree (solid squares) in Fig.~\ref{fig4}.
We also show as the open symbols the counts computed
using the minimum (triangles) and maximum (diamonds) amplifications
of Table~3. The shape and the normalization of the counts are quite 
insensitive to the choice of minimum or maximum amplifications 
(see also \markcite{chapman02}Chapman et al.\ 2002).
However, the flux to which the counts can be considered
to be determined does depend critically on the amplification. 

In the present data, if we adopt the minimum amplifications, then
the counts only extend down to 0.1~mJy. However, if we were instead to
assume that the average or maximum amplifications were appropriate, 
then the counts would extend to much fainter fluxes ($<0.1$~mJy). 
Because the areas of high amplification are small, this would then
imply that the counts turn up quite steeply below 0.1~mJy 
in order for us to actually be able to detect sources in such small 
areas. Indeed, in the maximum amplification case, two of the 
sources even lie below the 0.01~mJy flux limit of Fig.~\ref{fig4}.
However, since we have a perfectly adequate solution with the
minimum amplifications, at present we have no way of knowing 
whether there really are sources at fluxes below 0.1~mJy. 
To avoid this uncertainty, for the remaining discussion of the 
counts we shall restrict ourselves to fluxes above 0.1~mJy.

We have overplotted on the figure a power-law fit (solid line) to 
the solid squares based on an area-weighted maximum likelihood fit
(\markcite{crawford70}Crawford, Jauncey, \& Murdoch 1970),
which gives a slope of $-1.1$ and provides a good representation
of the raw counts.
Given the complex effects of noise and confusion, the errors 
and systematic biases of the counts are best estimated with 
Monte Carlo simulations. 

\section{Simulations}
\label{secsms}

We created simulated images by drawing sources from a population 
with a count described by the power-law fit and
placing them randomly on the source plane. We 
limited the fluxes of the input sources to be brighter than 0.01~mJy. 
We then mapped and amplified the sources to the image 
plane using {\sc LENSTOOL} and added the sources into the true 
noise maps. We analyzed the resulting images with the same 
procedures that we used to analyze the real data. We ran 100 realizations
for each field and derived the average output counts. We then
iteratively adjusted the input count model until the average output
counts matched the observed counts. For the final determination 
of the counts completeness and confidence
ranges we used the input power-law
\smallskip

\begin{equation}
N(>S)=3.5\times 10^3\times (S/2~{\rm mJy})^{-1.2}~{\rm deg}^{-2}
\end{equation}

\smallskip
\noindent
shown by the dashed line in Fig.~\ref{fig5}.
The solid line in Fig.~\ref{fig5} shows the recovered counts obtained 
for this input count model. 
The recovered counts can be compared with the observed raw counts, 
which are shown by the solid squares in Fig.~\ref{fig5}. The recovered 
counts exhibit a small systematic upward (Eddington) bias relative 
to the input counts at the brighter fluxes.
Over the 1 to 5~mJy range we see an increase by a factor of 1.25.
\markcite{eales00}Eales et al.\ (2000) and 
\markcite{scott02}Scott et al.\ (2002) obtained a similar 
systematic flux boost from their simulations. Eales et al.\
found a median boost factor of 1.44 for their survey
down to 3~mJy, and Scott et al.\ found boost factors of 1.28
and 1.35 at 5~mJy for their two areas. The shallower slope
of the counts in the present simulation may account for the
slightly smaller correction. At low fluxes the recovered
counts drop below the input counts as the source detection
becomes incomplete. 

The uncertainty corridors for the counts may also be obtained by
looking at the spread of counts over the various realizations.
The 90\% confidence range measured in this way from the
simulations is shown by the thin solid lines and should provide
a good estimate of the uncertainties on the observed counts.

%
% Figure 5
%

\begin{inlinefigure}
\psfig{figure=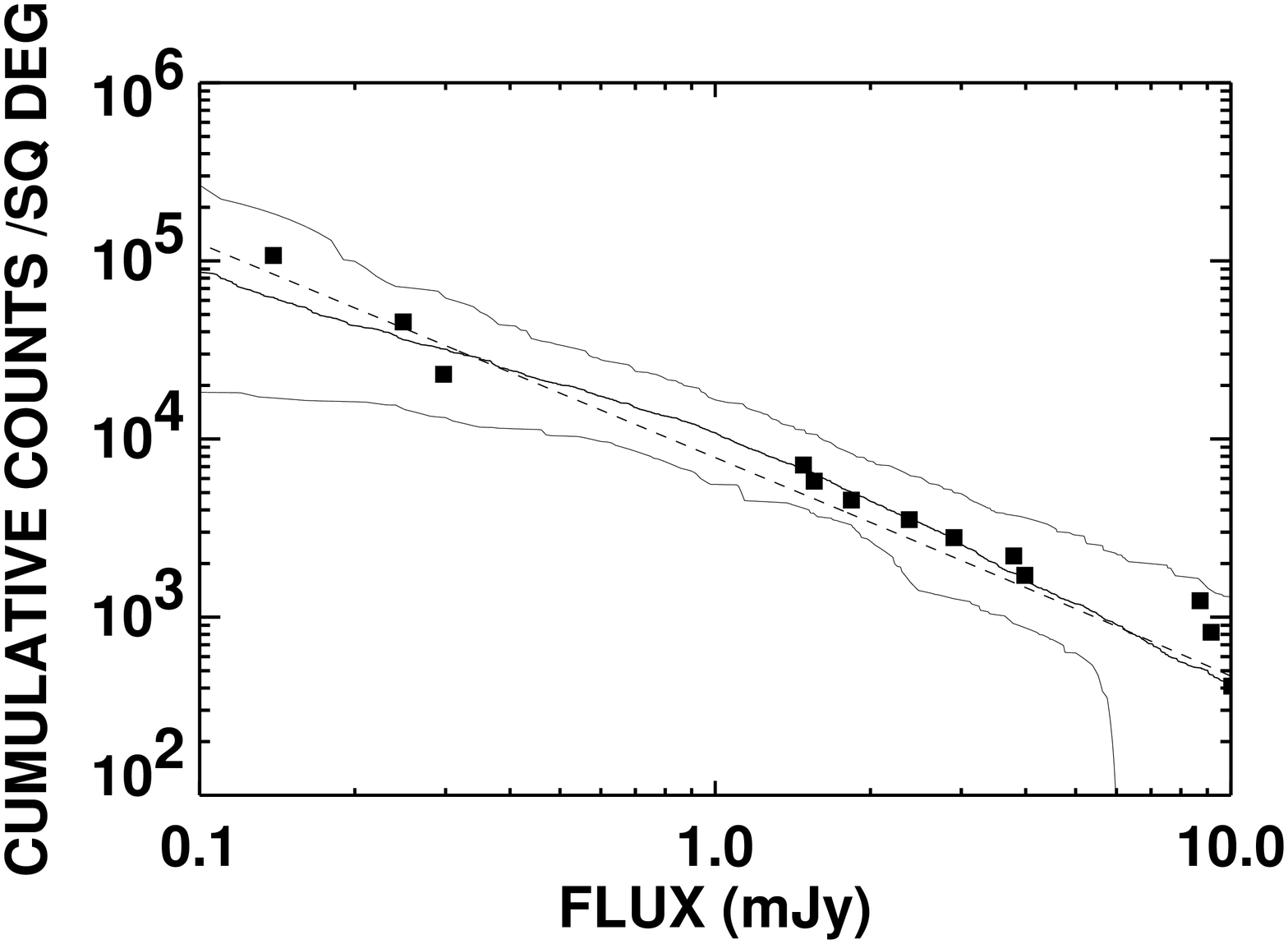,angle=90,width=3.5in}
\vspace{6pt}
\figurenum{5}
\caption{
Comparison of the observed counts (solid boxes) with the
simulations. The solid
curve shows the returned counts from the Monte Carlo simulations,
which were based on an input model described by a power-law fit
(dashed line). Eddington bias
raises the measured counts above the input counts at the
high flux end, while at the low end incompleteness in the
source detection drops the measured counts below the input
counts.
The thin solid lines show the 90\% confidence range calculated
using the simulations.
\label{fig5}
}
\addtolength{\baselineskip}{10pt}
\end{inlinefigure}

As is evident from Fig.~\ref{fig5}, the counts are best determined 
in the 0.3 to 5~mJy flux
range. Here the areas are large enough to provide a significant
sample. However, the statistical uncertainties are still large
with only 7 sources lying in this range. 
At 1~mJy we find a cumulative count of 
$0.9_{0.5}^{1.4}\times 10^4$~deg$^{-2}$, after correcting for 
the systematic upward bias, where the upper and
lower bounds are the 90\% confidence range. This is consistent 
with the \markcite{blain99}Blain et al.\ (1999) measurement
of $(0.79\pm 0.3)\times 10^4$~deg$^{-2}$.
At 0.3~mJy the cumulative counts have risen to 
$3.3_{1.3}^{6.3}\times 10^4$~deg$^{-2}$, where the upper and
lower bounds are again the 90\% confidence range.

\section{Noise Analysis}
\label{secnoise}

The noise distributions of the uncleaned noise maps (i.e., the maps
with the directly detected sources removed) contain additional
independent information over the direct counts.
In principle, the noise distributions should provide a more
sensitive diagnostic of the counts at sub-mJy fluxes than the
direct counts because the useable areas are much larger than the
areas over which $3\sigma$ sources can be detected.

In Fig.~\ref{fig6} we show the distribution functions (combined
for all three cluster fields) from both the true noise and the
uncleaned noise maps. The true noise distribution function
is well fit by a smooth Gaussian
function, which is shown in the figure in place of the actual
distribution function. The uncleaned noise distribution function
(jagged curve) has extensions to the right and left of the true noise
distribution function. The different shapes of these two extensions
result from there being twice as many negative positions as positive
positions in the beam pattern corresponding to each source, but
at half the flux (because of the
nod and fixed chop observing procedure used).
For the present data
a Kolmogorov-Smirnov test (applied to the absolute values of
the distribution to combine the two tails of the distribution
function) does not reject the hypothesis
that the uncleaned noise distribution is drawn from the 
true noise distribution because there are a relatively small number 
(approximately 240) of independent points in the fields.
A noise analysis is therefore primarily useful in placing an 
upper bound on the counts.

%
% Figure 6
%

\begin{inlinefigure}
\psfig{figure=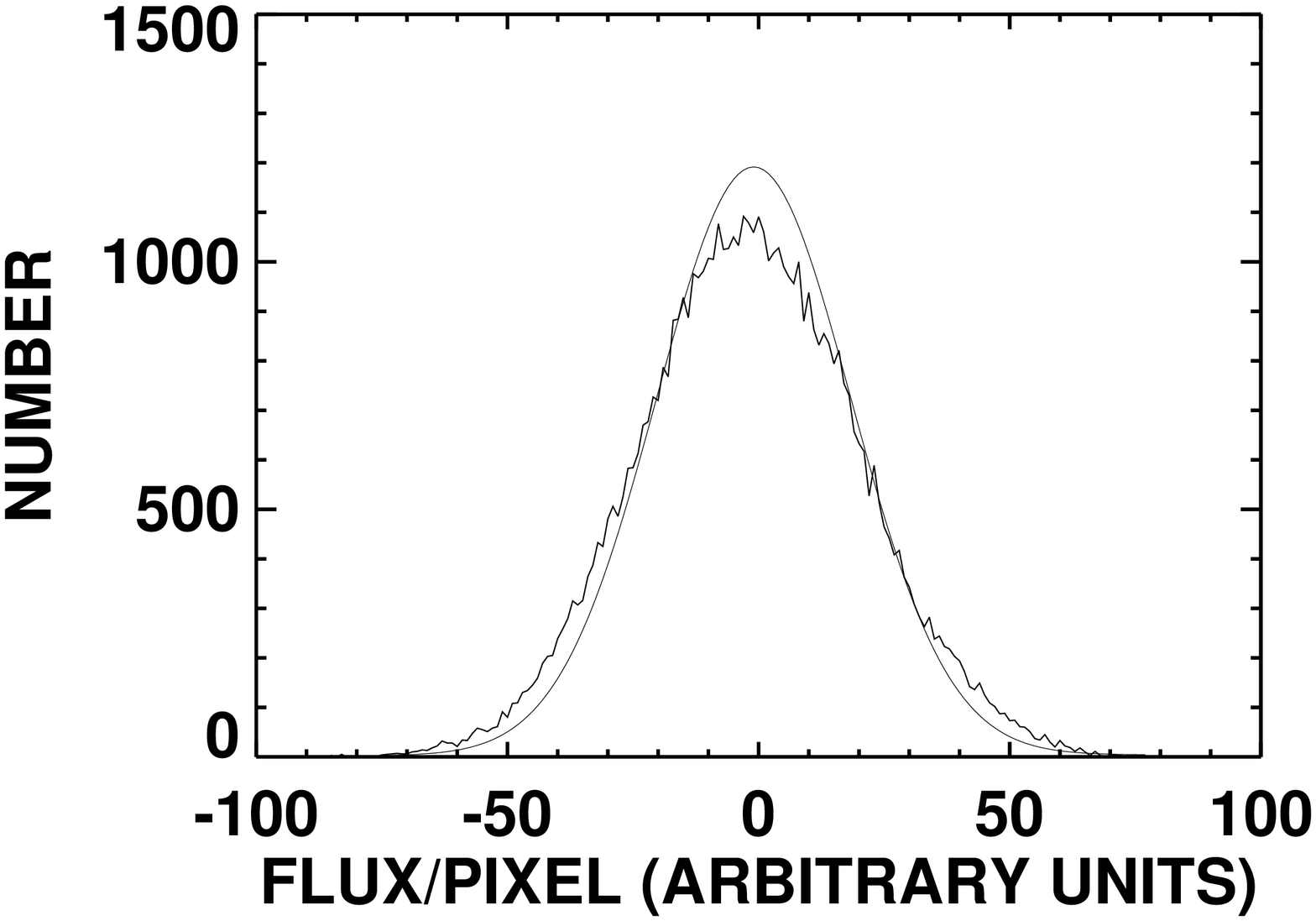,angle=90,width=3.5in}
\vspace{6pt}
\figurenum{6}
\caption{Distribution function for the three fields.
The true noise is well fit by a smooth Gaussian function
(solid curve). The extensions of the uncleaned noise
(jagged curve) indicate the presence of faint sources below the
detection limit. The differences between the right-hand and
left-hand extensions result from there being twice as many
negative positions as positive in the beam pattern corresponding
to a given source, but at half the flux.
\label{fig6}
}
\addtolength{\baselineskip}{10pt}
\end{inlinefigure}

The difference between the observed uncleaned and the simulated
uncleaned distribution functions can be used to constrain the
simulated counts distribution.
If the input function for the counts has too high a normalization
or is too steep at the faint end, then the simulated distribution
will be too wide when compared to the observed distribution.
In contrast, if the input function has too low a normalization or
is too shallow at the faint end, then the simulated distribution 
will be too narrow when compared to the observed distribution.

We performed Monte Carlo simulations using two different
parameterizations to determine the number
counts at faint fluxes. In the first case we used the
\markcite{bcs99}Barger et al.\ (1999a) parameterization

\begin{equation}
n(S)=N_o/(a+S^\alpha)
\end{equation}

\noindent
with $S$ in mJy, which fit their differential blank field
counts above 2~mJy for $\alpha = 3.2$ and
$N_o=3.0\times 10^4$~deg$^{-2}$~mJy$^{-1}$. Here we varied $a$.
In the second case we used a power-law parameterization
of the faint counts normalized to match the
Barger et al.\ counts and the present direct counts at 2~mJy.
Here we varied the power-law index.

Sources were drawn from a population with a count described by
the chosen model. For each of the two models we performed 100
realizations at each value of the index. For each realization
we randomly populated the source plane of each cluster, assuming
source redshifts of $z=3$. The source plane sources
were then mapped onto the image plane using the
{\sc LENSTOOL} model and added into the true noise map of the
field to construct the simulated image.

%
% Figure 7
%

\begin{inlinefigure}
\psfig{figure=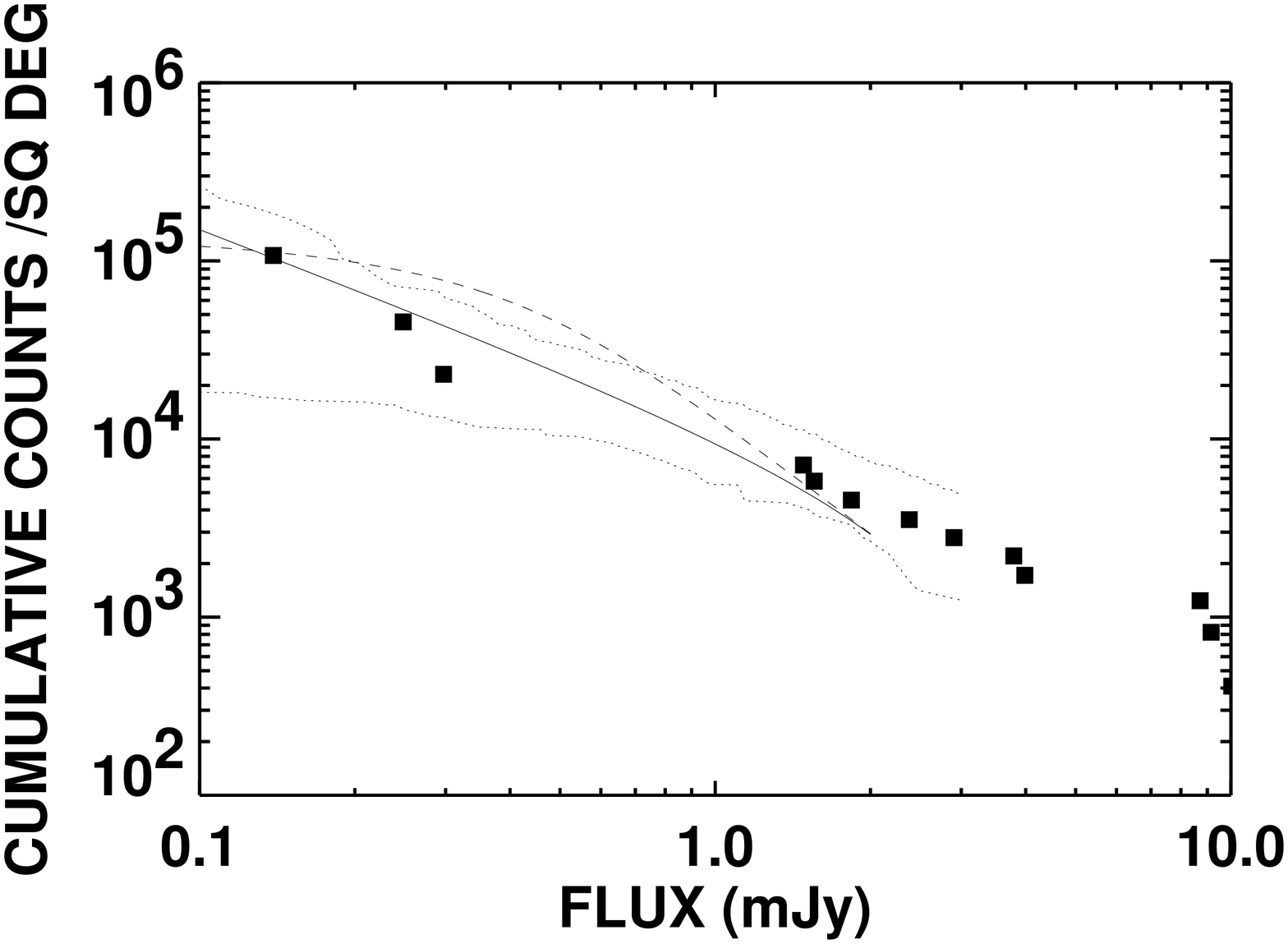,angle=90,width=3.5in}
\vspace{6pt}
\figurenum{7}
\caption{
The dashed curve shows the 90\% confidence upper limits from
Monte Carlo simulations that assumed the parameterization
of Barger et al.\ (1999a), $n(S)=N_o/(a+S^\alpha)$, with
$N_o=3.0\times 10^4$~deg$^{-2}$~mJy$^{-1}$ and $\alpha=3.2$;
the value of $a$ for the curve is 0.13.
The solid power-law curve shows the 90\% confidence limit
from Monte Carlo simulations that assumed a power-law
parameterization. The power-law index is 2.1.
The solid boxes show the direct counts and the jagged curves
show the 90\% confidence range on the direct counts.
\label{fig7}
}
\addtolength{\baselineskip}{10pt}
\end{inlinefigure}

We measured the simulated distribution functions in the same way
that we measured the uncleaned noise distribution function. That
is, we first removed all $>3\sigma$ sources in order to sample
the effect of the faint sources. We then applied a Kolmogorov-Smirnov
test to compare the simulated distributions with the observed 
uncleaned distribution. We found that at the 90\% confidence level
the index $a$ in the Barger et al.\ (1999a) parameterization cannot 
be less than 0.13, and the power-law index cannot exceed 2.1. 
These upper bounds are shown as the dashed line and solid 
power-law, respectively, in Fig.~\ref{fig7}. 
These independent estimates from the noise analysis closely 
match the upper limits from the direct counts over the flux range 
above 0.2~mJy but place a somewhat tighter
constraint at 0.1~mJy where the 90\% confidence limit on the cumulative
counts is $1.5\times 10^5~{\rm deg}^{-2}$ for the power-law model.

\section{Discussion}
\label{secdisc}

We summarize the current results and
compare them with wide-field blank surveys in Fig.~\ref{fig8}.
The present data are shown by the solid squares. We 
show the lensing analysis of 
\markcite{blain99}Blain et al.\ (1999) as open
circles and that of \markcite{chapman02}Chapman et al.\ (2002) 
as open downward pointing triangles. The present analysis  
shows good overlap with that of Blain et al.\ at the
brighter fluxes, though both are slightly lower than
that of Chapman et al. In order to show the counts at bright
fluxes we have plotted the blank field surveys of
Barger, Cowie, \& Sanders (1999), Hughes et al.\ (1998), 
Eales et al.\ (2000), Scott et al.\ (2002), and 
Borys et al.\ (2002). The large area survey of Scott et al.\
and the Hubble Deep Field scan map point of Borys et al.\ 
produce slightly shallower counts than the other surveys. 

We show a simple broken power-law fit to the counts in 
Fig.~\ref{fig8} where the differential counts are given by 

\begin{equation}
n(S)=N_0/S^{\alpha}
\end{equation}

\noindent
with $N_0=2.5\times 10^{4}$~deg$^{-2}$~mJy$^{-1}$ and $\alpha=3.0$
above $S=3$~mJy and $N_0=1\times 10^{4}$~deg$^{-2}$~mJy$^{-1}$ and 
$\alpha=2.2$ below 3~mJy. The EBL in this representation diverges 
at the faint end. We plot the Barger et al.\ (1999) parametric fit
for two values of $\alpha$ (dashed line shows 3.0, dotted line 
shows 3.2), $N_0=3\times 10^{4}$~deg$^{-2}$~mJy$^{-1}$, 
and $a=0.7$ and $a=0.5$, respectively.
Both curves give an integrated surface brightness set to 
match the EBL measurement of 
\markcite{fixsen98}Fixsen et al.\ (1998) and
provide reasonable descriptions of the
counts, though the shallower model may be preferred
as a better match to the \markcite{scott02}Scott et al.\ (2002) 
data.

%
% Figure 8
%

\begin{inlinefigure}
\psfig{figure=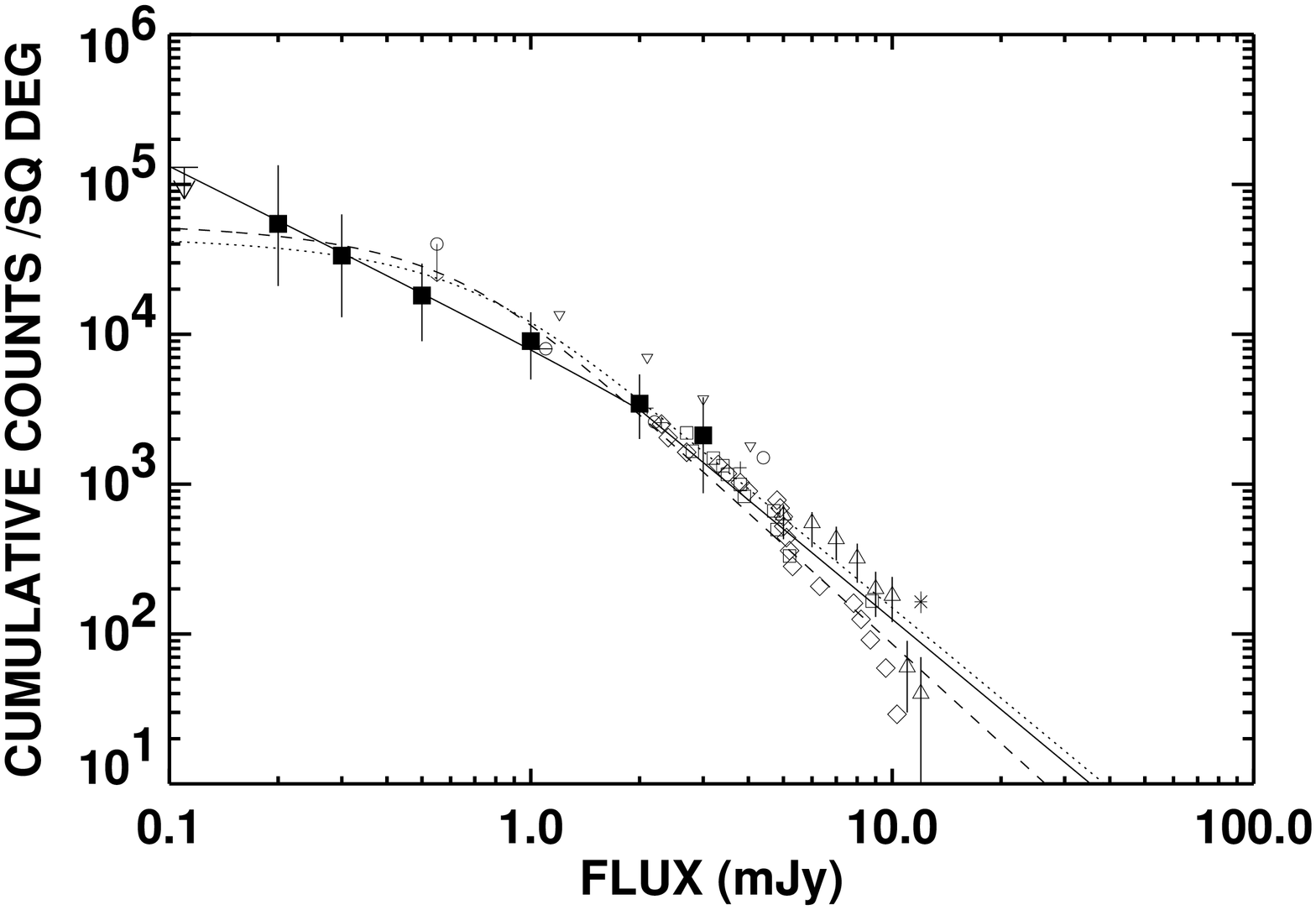,angle=90,width=3.5in}
\vspace{6pt}
\figurenum{8}
\caption{
A summary of the counts and 90\% confidence limits between
0.2 and 3~mJy from the analysis of \S~5 is shown by the solid 
squares and associated uncertainties.
The upper limit at 0.1~mJy, based on the noise analysis
of \S~6, is shown as a downward pointing arrow.
The open circles show the points from Blain et al.\ (1999).
These have been slightly displaced to larger fluxes
to distinguish them from the present counts.
The Blain et al.\ point at 0.5~mJy is shown
as a $2\sigma$ upper limit for consistency with the
present analysis. The open downward pointing triangles
show the cluster lensing analysis counts of Chapman et al.\ (2002).
All of the lensing analysis data is only shown below a flux of 5~mJy.
We include the wide-field counts of Barger, Cowie,
\& Sanders (1999) (open diamonds), Hughes et al.\ (1998)
(crosses), Eales et al.\ (2000) (open squares), 
Scott et al.\ (2002) (open triangles), and Borys et al.\ (2002)
(asterisk), but we only show
$1\sigma$ uncertainties for the Scott et al.\ data to
avoid confusing the plot; the uncertainties on the
other surveys are larger. We show the
Barger, Cowie, \& Sanders (1999) parametric fit
(dotted line) and an alternate version (dashed line) 
of this parametric fit 
discussed in the text, which provides a better match to the 
Scott et al.\ data; it also integrates to match the EBL. 
We also show a broken power-law representation of the data
with slope $-2$ above 3~mJy and $-1.2$ below 3~mJy. This
representation is divergent at faint fluxes and
must turn over further at some point.
\label{fig8}
}
\addtolength{\baselineskip}{10pt}
\end{inlinefigure}

The contribution to the EBL in the 0.3 to 2~mJy flux 
range can be directly measured from our counts and
is $2.0_{0.8}^{3.2}\times 10^4$~mJy~deg$^{-2}$,
where the upper and lower bounds are the 90\% confidence range.
Thus, the percentage of the EBL residing
in this range is $65_{26}^{100}$\%, if we adopt the
850~$\mu$m EBL measurement of $3.1\times 10^4$~mJy~deg$^{-2}$
from \markcite{puget96}Puget et al.\ (1996), or
$45_{18}^{72}$\%, if we adopt the measurement of
$4.4\times 10^4$~mJy~deg$^{-2}$ from
\markcite{fixsen98}Fixsen et al.\ (1998).
Given that $20-30$\% of the EBL is resolved at fluxes between 
2 and 10~mJy (\markcite{bcs99}Barger et al.\ 1999a; 
\markcite{eales99}Eales et al.\ 1999, 2001, 
\markcite{scott02}Scott et al. 2002), 
it appears that most of the submillimeter EBL is 
arising in sources brighter than 0.3~mJy.

\acknowledgements
We thank the referee for a very thorough review
which improved the manuscript. LLC gratefully acknowledges 
support from NSF through grant AST-0084816. 
AJB gratefully acknowledges
support from NASA through Hubble Fellowship grant HF-01117.01-A 
awarded by the Space Telescope Science Institute, which is 
operated by the Association of Universities for Research in 
Astronomy, Inc. for NASA under contract NAS 5-26555,
from NSF through grant AST-0084847, from the University of Wisconsin
Research Committee with funds granted by the Wisconsin Alumni 
Research Foundation, and from the American Association for 
University Women Educational Foundation and the American 
Astronomical Society through the Annie Jump Cannon Award.

\newpage

\newpage

\begin{deluxetable}{crrccccc}
\renewcommand\baselinestretch{1.0}
\tablewidth{0pt}
\tablecaption{SCUBA Observations of the A370, A851, and A2390 Fields}
\small
\tablehead{Cluster & RA(2000) & Dec(2000) & Total Exposure & 
Median & Area & True Noise & Uncleaned Noise \cr
%\multicolumn{2}{c}{$1\sigma$ (mJy)} \cr
%& & & (ks) & & (arcmin$^2$) & True & Uncleaned \cr
& & & Time (ks) & $\tau(850~\mu$m) & (arcmin$^2$) & 
($1\sigma$)(mJy) & ($1\sigma$)(mJy) \cr
}
\startdata
A370  &  2 39 53.10 & $-1$ 34 35.0 & 133.1 & 0.18 & 6.3 & 0.35 & 0.45 \cr
A851  &  9 42 59.65 & 46 58 57.0   & 59.3 & 0.27 & 6.3 & 0.80 & 0.87 \cr
A2390 & 21 53 35.55 & 17 41 52.3   & 78.1 & 0.26 & 6.3 & 0.69 & 0.71 \cr
\enddata
\label{tab1}
\end{deluxetable}

\begin{deluxetable}{crrrr}
\renewcommand\baselinestretch{1.0}
\tablewidth{0pt}
\tablecaption{Cluster Submillimeter Sources in the A370, 
A851, and A2390 SCUBA Fields}
\small
\tablehead{\# & RA(2000) & Dec(2000) & $f(850~\mu$m) & S/N \cr
& & & (mJy) & }
\startdata
C0 &  2 39 53.07 & $-1$ 34 55.7 & $ 1.97\pm 0.46$ &   4.34  \cr
C1 &  2 39 52.69 & $-1$ 34 18.6 & $-0.41\pm 0.46$ & $-0.90$ \cr
C2 &  9 42 56.14 &  46 59 12.5 & $ 0.36\pm 1.05$ &   0.35  \cr
C3 &  9 42 57.93 &  46 59 12.7 & $-1.93\pm 0.95$ & $-2.04$ \cr
C4 & 21 53 36.75 &  17 41 44.2 & $ 6.96\pm 0.67$ &  10.39  \cr
\enddata
\label{tab2}
\end{deluxetable}

\begin{deluxetable}{crrrrcccrc}
\renewcommand\baselinestretch{1.0}
\tablewidth{0pt}
\tablecaption{Background Submillimeter Sources in the A370, 
A851, and A2390 SCUBA Fields} 
\small
\tablehead{\# & RA(2000) & Dec(2000) & $f(850~\mu$m) & S/N & Amplification
& $z$ & Smail et al. & RA Offset & Dec Offset \cr
& & (mJy) & & & (min, max) & & $f(850~\mu$m)(mJy) & (sec) & (arcsec)}
\startdata
0 & 2 39 51.90   & $-1$ 35 59.0 & $21.06\pm 1.34$ & 15.73 &  2.3      &
2.80 & 23.0 & $-0.00$ & 0.0 \cr

1 & 2 39 56.63   & $-1$ 34 27.0 & $ 6.68\pm 0.58$ & 11.58 &  2.3 &
1.06 & 11.0 & $-0.23$ & 0.0 \cr

2 & 2 39 57.64   & $-1$ 34 56.0 & $ 3.49\pm 0.66$ &  5.29 &  1.9 (1.8,2.0) &
\nodata & \nodata & \nodata & \nodata \cr

3 & 2 39 58.57   & $-1$ 34 34.0 & $ 2.52\pm 0.74$ &  3.40 &  1.7 (1.6,1.7) &

\nodata & \nodata & \nodata & \nodata \cr

4 & 2 39 53.83   & $-1$ 33 37.0 & $ 2.17\pm 0.57$ &  3.82 &  7.3 (5.6,10.8) &
\nodata & \nodata & \nodata & \nodata \cr

5 & 2 39 52.63   & $-1$ 34 40.0 & $ 1.39\pm 0.45$ &  3.09 & 10 (4.2,43) &
\nodata & \nodata & \nodata & \nodata \cr

6 & 2 39 47.36   & $-1$ 35 07.0 & $ 2.49\pm 0.72$ &  3.43 &  1.6 (1.6,1.7)&
\nodata & \nodata & \nodata & \nodata \cr

7 & 9 42 54.57   &  46 58 44.0 & $15.06\pm 1.07$  & 14.02  &  1.5 (1.5,1.6) &
\nodata & 17.2 & 0.13 & 0.0 \cr

8 & 9 43 03.96   &  47 00 16.0 & $10.46\pm 1.83$  & 5.72   &  1.2 (1.2,1.2) &
\nodata & \nodata & \nodata & \nodata \cr

9 & 9 42 53.49   &  46 59 52.0 & $4.93\pm 1.49$  & 3.30   &  1.3 (1.3,1.4) &
\nodata & \nodata & \nodata & \nodata \cr

10 & 21 53 33.31 &  17 42 49.3 & $ 7.57\pm 0.93$  & 8.14   &  1.9 (1.8,2.0) &
\nodata & \nodata & \nodata & \nodata \cr

11 & 21 53 38.35 &  17 42 16.3 & $ 4.52\pm 1.04$  & 4.35   &  1.9 &
1.02 & 6.7 & 0.15 & 3.3 \cr

12 & 21 53 35.48 &  17 41 09.3 & $3.24\pm 0.78$   & 4.15   &  52 (0.6,52) &
\nodata & \nodata & \nodata & \nodata \cr

13 & 21 53 38.21 &  17 41 52.3 & $2.64\pm 0.72$   & 3.64   &  11 ($>6.7$)&
\nodata & \nodata & \nodata & \nodata \cr

14 & 21 53 34.15 &  17 42 02.3 & $2.64\pm 0.72$   & 3.42   &  39 ($>15$)&
\nodata & \nodata & \nodata & \nodata \cr
\enddata
\label{tab3}
\end{deluxetable}

\end{document}